\documentclass[conference]{IEEEtran}
\IEEEoverridecommandlockouts

\usepackage[left=1.62cm,right=1.62cm,top=1.9cm]{geometry}
\usepackage{cite}
\usepackage{amsmath,amssymb,amsfonts}
\usepackage{algorithmic}
\usepackage{graphicx}
\usepackage{textcomp}
\usepackage{amsmath}
\DeclareMathOperator{\diag}{diag}
\usepackage{amsmath}
\usepackage{mathtools}
\usepackage{verbatim}
\usepackage{xcolor}

\usepackage[T1]{fontenc}
\usepackage[utf8]{inputenc}
\usepackage{authblk}

\def\BibTeX{{\rm B\kern-.05em{\sc i\kern-.025em b}\kern-.08em
    T\kern-.1667em\lower.7ex\hbox{E}\kern-.125emX}}
\begin{document}

\title{Impact of RIS on Outage Probability and Ergodic Rate in Wireless Powered Communication\\

{
} 
\thanks{This work was supported by Basic Science Research Program through NRF
funded by the MOE (NRF-2022R1I1A1A01071807, 2021R1I1A3041887), and
Institute of Information \& communications Technology Planning \& Evaluation
(IITP) grant funded by the Korea government (MSIT) (2022-0-00704, and Development
of 3D-NET Core Technology for High-Mobility Vehicular Service).}
}

\author[*]{Waqas Khalid}
\author[$\S$]{Manish Nair}
\author[$\ddagger$]{Trinh Van Chien}
\author[$\dagger$]{Heejung Yu}
\affil[*]{Institute of Industrial Technology, Korea University, Sejong, Korea; waqas283@\{korea.ac.kr, gmail.com\}}
\affil[$\S$]{CSN Research Group, University of Bristol, Bristol, U.K; manish.nair@bristol.ac.uk}
\affil[$\ddagger$]{Hanoi University of Science and Technology, Hanoi, Vietnam; chientv@soict.hust.edu.vn}
\affil[$\dagger$]{Dept. of Elec. \& Inform. Eng., Korea University, Sejong, Korea; heejungyu@korea.ac.kr}
\renewcommand\Authands{ and }




\maketitle

\begin{abstract}

Wireless powered communication (WPC) combines information and energy transmission for energy-constrained nodes. Reconfigurable intelligent surfaces (RISs) are capable of controlling radio signals in a dynamic and goal-oriented manner. This paper investigates the combination of RIS and WPC to enhance the performance of an energy-constrained user. Using a RIS, a base station, and a wireless user transmit energy and information signals, respectively. We derive closed-form expressions for outage probability and secrecy rate to analyze the performance of the proposed framework. Based on the theoretical analysis and simulation results, valuable insights are revealed and parameter selection is demonstrated.
\end{abstract}


\section{Introduction}
Future sixth-generation (6G) networks will utilize a reconfigurable intelligent surface (RIS) to achieve a smart wireless environment and address issues related to radio-frequency (RF) impairment and signal propagation. Wireless connectivity under harsh propagation conditions is enhanced via RIS through non-line-of-sight (nLoS) links and controlled scattering and multipath components. A RIS offers better coverage and reliability without sophisticated signal processing or RF operations. In particular, RIS systems are composed of a large number of passive reflective elements that allow phase shifts to be introduced into electromagnetic (EM) signals without deploying additional base stations (BSs) or relay nodes. A RIS can enable the signals to be constructively combined to enhance the received signal at the intended receiver to support ubiquitous communications and diverse user requirements. In terms of design and implementation, passive RIS has attractive characteristics, such as low power consumption, low hardware costs, and high flexibility \cite{b1,b2}.

In other developments, simultaneous wireless information and power transfer (SWIPT) and wireless powered communication (WPC) enable the transmission of both information and energy \cite{b2,b3,b4,b5}. In particular, a device in SWIPT simultaneously decodes information and harvests energy from received RF signals \cite{b2,b3}, while WPC node performs downlink energy harvesting before uplink information transmission \cite{b4,b5}. WPC is a promising technological advancement that can significantly enhance the battery lifespan of energy-constrained wireless devices. Furthermore, RIS and WPC can be combined to further improve the performance and energy efficiency.

\subsection{Motivation and Contribution}
In prior research work, a RIS-WPC network has not been investigated where RIS is exploited to improve both energy transfer and information transmission. To this end, we aim to exploit the potential beamforming gain of RIS to enhance both energy transfer and information transmission. In particular, we propose a general framework for RIS-WPC network in which an energy-constrained wireless user ($U$) harvests RF energy from the BS and then uses the harvested energy to transmit information to a BS. Since the deployment of RIS to simultaneously enhance the energy transfer of $U$ and information reception of BS offers more adjustable parameters, previous results of WPC without RIS cannot be applied. We present a statistical characterization of the channels for tractable analysis and derive the close-form expressions of the ergodic rate and outage probability over Rayleigh fading channels. Using numerical results, we confirm the theoretical analysis and demonstrate the choice of parameters for the system.

\begin{figure}[t]
\centering
\includegraphics[width=3.4in,height=2.3in]{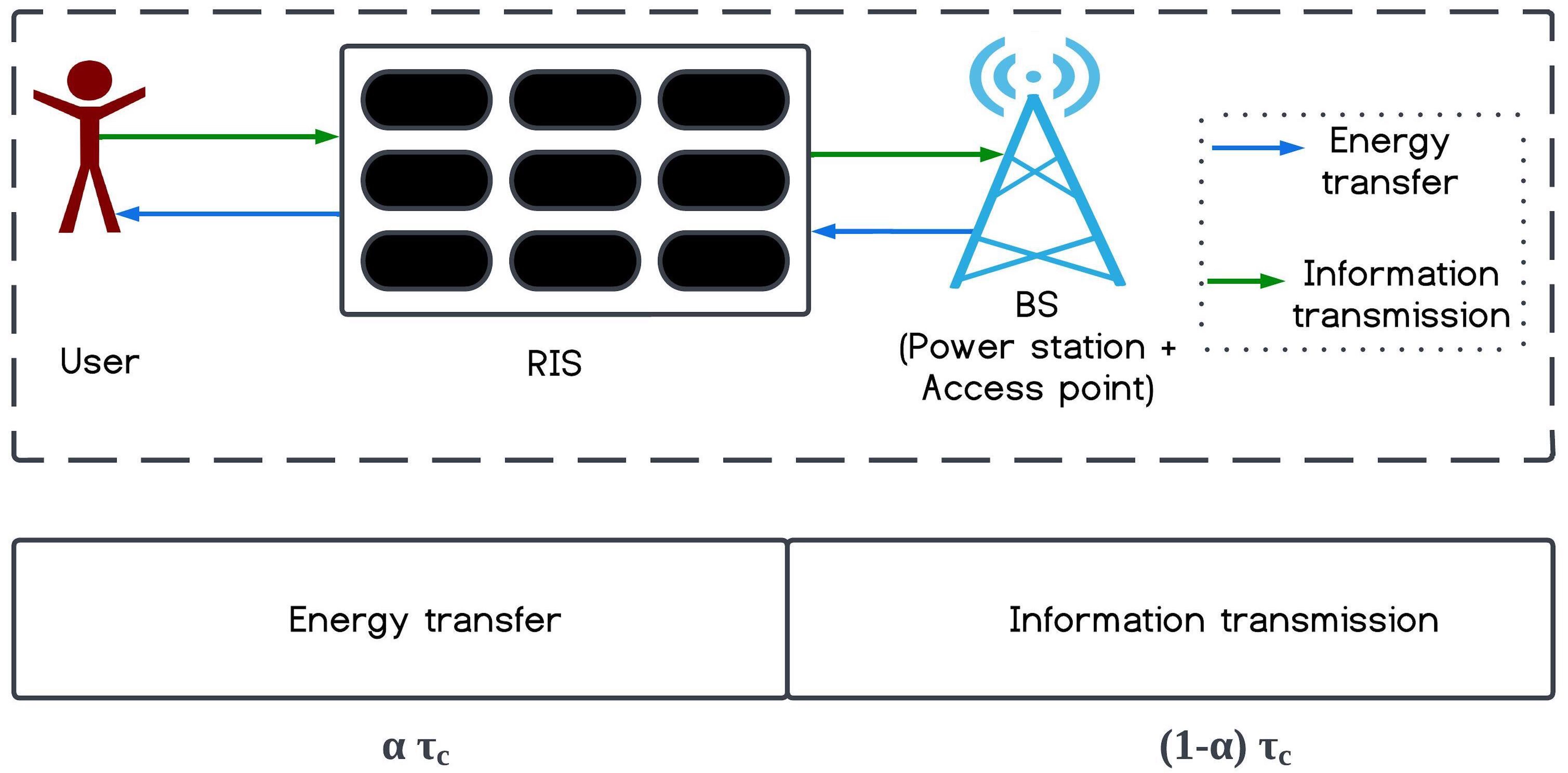}
\caption {RIS-WPC network with a generalized transmission frame structure.}
\label{diagram1}
\end{figure}

\section{System Model}

$\textit{Network Detail:}$ Consider a typical WPC network consisting of a BS and an energy-constrained user ($U$) as shown in Fig. \ref{diagram1}. For simplicity, a single antenna scenario is considered for all nodes\footnote{Our future work will extend to multiple antennas, where beamforming can be used at nodes with multiple antennas to improve system performance.}. A RIS consisting of $M$ elements is deployed between a BS and an $U$ to provide amplitude and phase responses for enhancing energy transfer and information transmission\footnote{We assume no direct link between BS and U due to an obstacle, which motivates the deployment of RIS to establish communication links.}.

$\textit{Channel Modeling:}$ All links are considered using the independent and identically distributed (i.i.d.) Rayleigh fading model and are reciprocal with quasi-static block fading, where the channel vectors remain the same within the channel coherence interval \cite{b6,b7}. The channel vectors between BS and RIS is $\textbf{h}_{br}$ and between $U$ and RIS is $\textbf{h}_{ur}$. Each entry of $\textbf{h}_{br}$ and $\textbf{h}_{ur}$ is assumed to be an independent complex Gaussian distribution with zero mean and unit variance.

$\textit{Signal Description:}$ Communication takes place in two phases during each channel coherence interval ($\tau_c$), namely energy transfer and information transmission. During $\alpha \tau_c$, BS transfers RF energy to $U$ via RIS. The total energy harvested by $U$ for a linear energy harvesting model can be written as,

 \begin{align}
 E=\eta  \alpha \tau_c P_{b}  |\textbf{h}_{ur}^T \mathbf{\Theta} \textbf{h}_{br} |^2 \zeta
  \end{align}
where $\mathbf{\Theta}\!=\!\diag  (\sqrt{\beta_1}e^{j\theta_1}, ..., \sqrt{\beta_{M}}e^{j\theta_{M}})$ denotes $M \times M$ diagonal response matrix of RIS, $\eta$ denotes the energy conversion efficiency, $P_{b}$ denotes the transmission power of BS, and $\zeta$ denotes the path loss from BS to $U$ via $m^{th}$ element of RIS.

Following this, $U$ transmits its information to BS via RIS during $(1-\alpha)\tau_c$ using the energy harvested during $\alpha\tau_c$. The transmission power of $U$ can be calculated as,

 \begin{align}
 P_u= \frac{E}{(1-\alpha)\tau_c}=\frac{\eta  \alpha  P_{b}  |\textbf{h}_{ur}^T \mathbf{\Theta} \textbf{h}_{br} |^2 \zeta}{(1-\alpha)}=P_e|\textbf{h}_{ur}^T \mathbf{\Theta} \textbf{h}_{br} |^2 \zeta,
  \end{align}
  where $P_e=\frac{\eta  \alpha  P_{b}}{(1-\alpha)}$

  At BS, the signal-to-noise ratio (SNR) can be written as,

    \begin{align}
   z_{u}=\frac{ P_e \zeta^2 |\textbf{h}_{ur}^T \mathbf{\Theta} \textbf{h}_{br} |^2 |\textbf{h}_{ur}^T \mathbf{\Theta} \textbf{h}_{br} |^2}{\sigma^2}=p_e \zeta^2 |\textbf{h}_{ur}^T \mathbf{\Theta} \textbf{h}_{br} |^2 |\textbf{h}_{ur}^T \mathbf{\Theta} \textbf{h}_{br} |^2
    \end{align}
  where $p_e=\frac{P_e}{\sigma^2}$, and $\sigma^2$ is variance of the additive white Gaussian noise.
  
$\textit{Phase Shift Design for RIS:}$ 
It is possible to increase the system's reliability through the beamforming gain provided by RIS. Channel state information (CSI) of all channels is assumed to be known, which can be obtained with some channel estimation schemes in RIS-aided communications \cite{b8}. To simplify the analysis, we assume continuous phase adjustments at the RIS. We also set the amplitude response to be maximum (i.e., independent of the phase-shift response). This implies that the phase shifts at the RIS can be intelligently controlled (can be matched with the phases of RIS-cascaded channels), i.e.,  $\theta_m\!=\!\arg \left(\textbf{h}_{ur,m}  \textbf{h}_{br,m} \right)$, to maximize the harvested energy of $U$ and the reception quality of BS. In this case, the SNR in Eq. (3) can be rewritten as,

\begin{align}
z_{u}=p_e \zeta^2 \left(\sum_{m=1}^{M}|\textbf{h}_{br,m}||\textbf{h}_{ur,m} |\right)^4
\end{align} 

\section{Performance Analysis}
The closed-form expressions of outage probability and ergodic rates are derived to analyze the performance of the proposed framework and obtain valuable insights.

\subsection{Outage Probability}
The outage probability is the probability of an event occurring when channel capacity drops below the codeword rate of the transmission signal ($r$), and is mathematically expressed as:

\begin{align}
P_{Out}=&\Pr\left[ \left(1-\alpha\right)\log_2\left(1+z_{u}\right)<r \right] \nonumber\\
=&\Pr\left[ \left(1-\alpha\right)\log_2\left(1+p_e \zeta^2 \left(\sum_{m=1}^{M}|\textbf{h}_{br,m}||\textbf{h}_{ur,m} |\right)^4\right)<r\right]
\end{align}

We first derive the statistic of $T=\sum_{m=1}^{M}|\textbf{h}_{br,m}||\textbf{h}_{ur,m} |$. $T$ can be approximated by a Gamma distribution\footnote{Due to the freedom to tune its parameters, the regular Gamma distribution is commonly used to approximate complicated distributions \cite{b1}.} with the shape parameter $k$ and scale parameter $w$. Using the moment matching technique, the parameters can be determined as,

\begin{align}
k=M\frac{\pi^2}{16-\pi^2},\; and \;\; \; w=\frac{16-\pi^2}{4\pi}
\end{align}

After some variable transformation, an approximation for the probability density function (PDF) of $T$ can be expressed as,

\begin{align}
f_T(t)=\frac{t^{k-1}}{\Gamma(k)w^{k}}e^{-\frac{t}{w}},
\end{align}
where $\Gamma(.)$ represents the Gamma function [9, Eq. (8.31)].

Using Eq. (7) and [9, Eq. (3.381.1)], we find out the outage probability as,

\begin{align}
P_{Out}\;=&\Pr \left[   T^4 < \frac{2^{\frac{r}{1-\alpha}}-1}{p_e \zeta^2}  \right] \nonumber\\ 
=& \int_{0}^{\sqrt[4]{\frac{2^{\frac{r}{1-\alpha}}-1}{p_e \zeta^2} }} f_T(t)dt \;= \; \frac{\gamma\left(k,\frac{1}{w}\frac{2^{\frac{r}{1-\alpha}}-1}{p_e \zeta^2}\right)}{\Gamma(k)}
\end{align}
where  $\gamma(.,.)$ denotes the lower incomplete Gamma function [9, Eq. (8.35)].

\subsection{Ergodic rate}
The ergodic rate of $U$ can be determined as \cite{b1},

\begin{align}
R_u=&\mathbb{E}\left[\log_2\left(1+p_e \zeta^2 \left(\sum_{m=1}^{M}|\textbf{h}_{br,m}||\textbf{h}_{ur,m} |\right)^4\right)\right]\nonumber\\ 
\approx& \log_2\left(1+p_e \zeta^2 \mathbb{E}\left[ \left(\sum_{m=1}^{M}|\textbf{h}_{br,m}||\textbf{h}_{ur,m} |\right)^4  \right] \right) 
\end{align}
where $\mathbb{E}\left[ \left(\sum_{m=1}^{M}|\textbf{h}_{br,m}||\textbf{h}_{ur,m} |\right)^4  \right]$ can be written as $\left(\frac{M\pi}{4}\right)^4 + 6\left(\frac{M\pi}{4}\right)^2 M\left(1-\frac{\pi^2}{16}\right)+3 \left(M\left(1-\frac{\pi^2}{16}\right)\right)^2$.

\section{Numerical Results}

Numerical results validate the theoretical expressions and provide valuable insights. Monte Carlo simulations with $10^6$ independent trials are used. Unless otherwise specified, the simulation parameters are as follows: $\alpha=0.4$, $\tau_c=1$, $\eta=0.85$, $r=1.2$ bps/Hz, $P_b=\{10,35\}$ dBm, and $M=50$. 

Fig. 2 shows the outage probability vs. number of RIS elements. The analytical values are calculated by Eq. (8). The results validate the analytical analysis that outage probability using RIS is determined by four powers. It implies that RIS has a significant impact and reliability can be enhanced significantly by increasing RIS elements. The reason is that RIS can simultaneously enhance energy transfer for $U$ and information reception for BS. Fig. 3 shows the ergodic rate vs. transmission power of BS (during energy transfer phase). The analytical values are calculated by Eq. (9).

\begin{figure}[t]
\centering
\includegraphics[width=2.6in,height=2in]{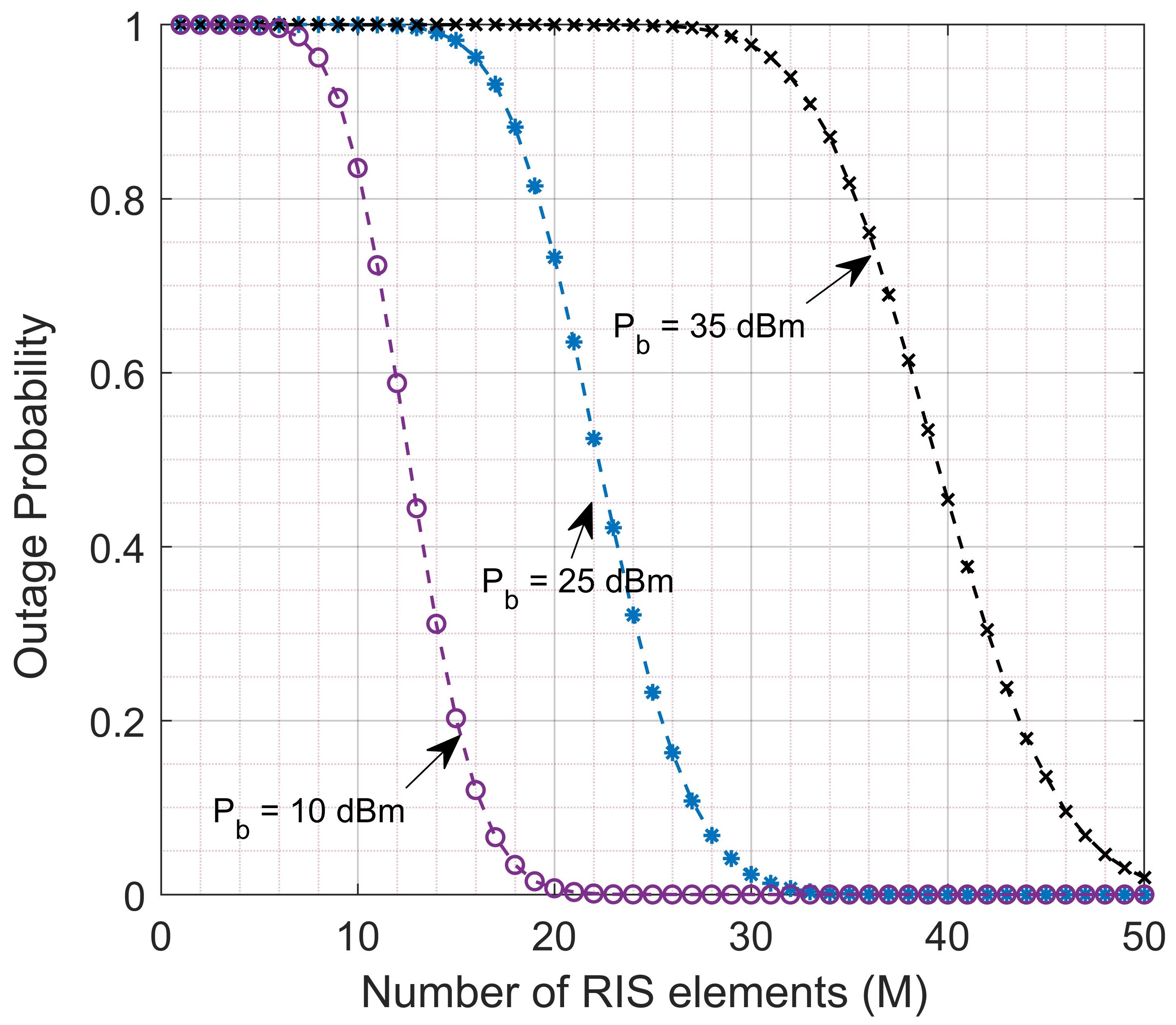}
\caption {Outage probability vs. Number of RIS elements ($M$) for different transmission power of BS ($P_b$)}
\label{diagram2}
\end{figure}

\begin{figure}[t]
\centering
\includegraphics[width=2.6in,height=2in]{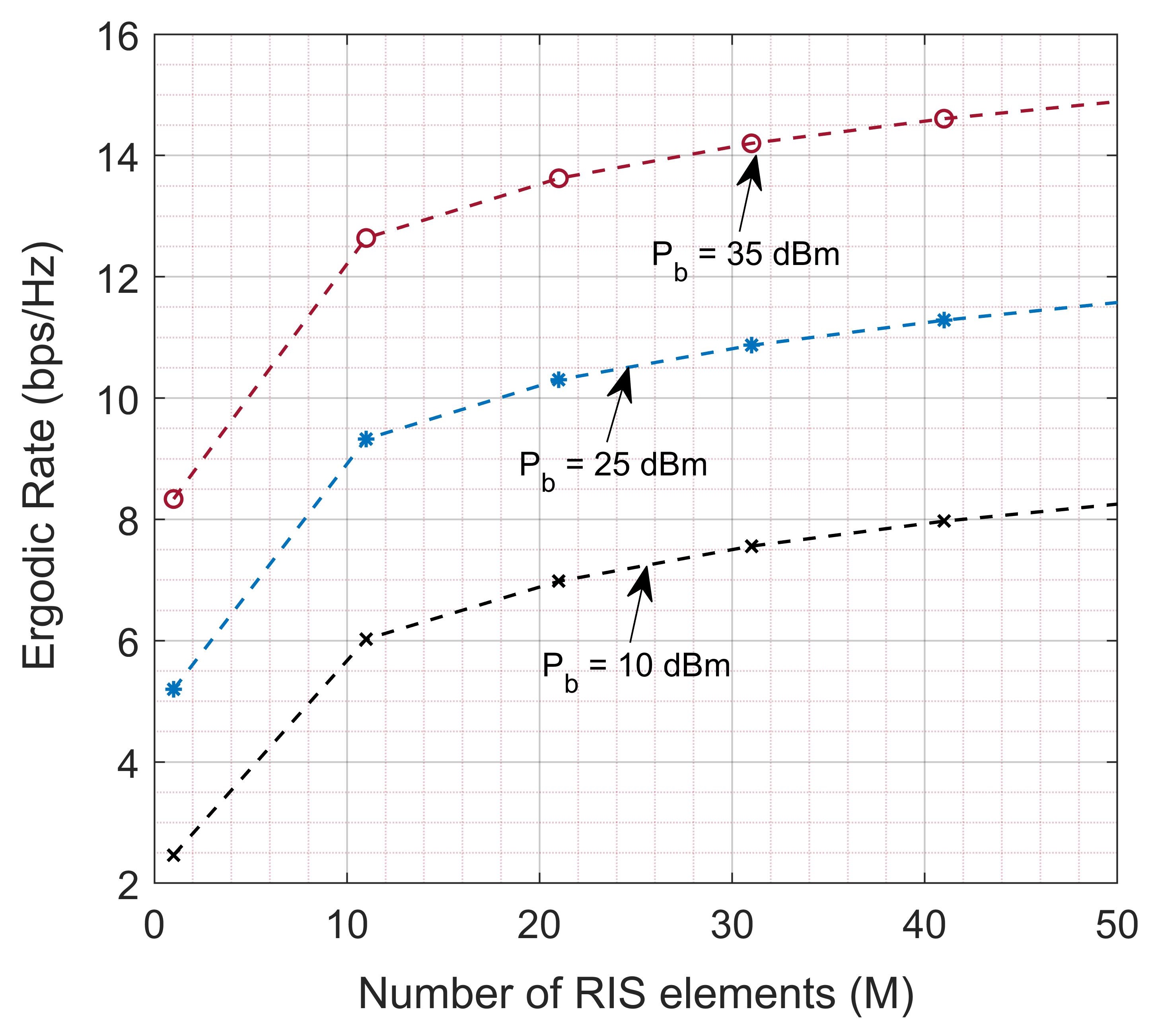}
\caption {Ergodic rate vs. Number of RIS elements ($M$) for different transmission power of BS ($P_b$)}
\label{diagram3}
\end{figure}

\section{Conclusion}
The combination of RIS and WPC enhances energy-constrained user performance.  In RIS-WPC networks, an energy-constrained user harvests RF energy from the BS and transmits information using the harvested energy. RIS is deployed to enhance user energy transfer and BS information reception. The proposed framework is evaluated using closed-form expressions of outage probability and secrecy rate. For outage probability, we use Gamma approximations for the RIS cascaded channels. The theoretical analysis and numerical results reveal valuable insights and illustrate parameter selection.

%



\end{document}